\documentclass[aps,prl,twocolumn,epsfig,amsmath,amssymb]{revtex4}
\usepackage[english]{babel} \usepackage{latexsym}
\usepackage{graphicx} \usepackage{subfigure} \usepackage{epsfig}
\usepackage{psfrag}
\usepackage{subfigure}
 
\begin{document}

\title{Phonon instability in two-dimensional dipolar Bose-Einstein Condensates}
\author{R. Nath$^1$, P. Pedri$^2$ and L. Santos$^1$} 
\affiliation{
\mbox{$^1$Institut f\"ur Theoretische Physik , Leibniz Universit\"at
Hannover, Appelstr. 2, D-30167, Hannover, Germany}\\
\mbox{$^2$Laboratoire de Physique Th\'eorique de la Mati\`ere Condens\'ee,
Universit\'e Pierre at Marie Curie,}\\
\mbox{case courier 121, 4 place Jussieu, 75252 Paris Cedex, France}\\
}

%
%
\begin{abstract}  

The partially attractive character of the dipole-dipole interaction
leads to phonon instability in dipolar condensates, which is followed by 
collapse in three-dimensional geometries. We show that the nature of this
instability is fundamentally different in two-dimensional condensates, due to
the dipole-induced stabilization of two-dimensional bright solitons. As a
consequence, a transient gas of attractive solitons is formed, and collapse
may be avoided. In the presence of an harmonic confinement, 
the instability leads to transient pattern formation followed by the 
creation of stable two-dimensional solitons. This dynamics should 
be observable in on-going experiments, allowing for the creation of 
stable two-dimensional solitons for the first time ever in quantum gases.

\end{abstract}  
 \maketitle


Recent experiments on atoms with large magnetic moments \cite{Pfau,
Stabilization,CollapsePfau,ParisNord},
polar molecules \cite{Jin}, and spinor Bose-Einstein condensates (BECs)
\cite{StamperKurn} are attracting a major interest to the physics of 
dipolar gases. In
these gases, the dipole-dipole interaction (DDI) plays a significant or even
dominant role when compared to the short-range isotropic 
interactions, which up to now
have played the major role in the physics of ultra cold gases. The DDI are
long-range and anisotropic, being partially attractive. As a consequence,
dipolar gases present a rich new physics for Bose gases
\cite{Santos2000,Roton,DipBEC}, Fermi gases \cite{DipFermi}, 
spinor gases \cite{DipSpinor}, strongly-correlated gases 
\cite{DipStrong}, and quantum information \cite{DipQInf}. 

BEC stability becomes an issue of obvious concern in the presence of
attractive interactions. 2D and 3D homogeneous short-range 
interacting condensates with attractive interactions 
($s$-wave scattering length $a<0$) are unstable against collapse. 
The presence of a trap may stabilize an attractive BEC 
for a sufficiently small number of particles \cite{Collapse}, whereas 
a sufficiently large interaction energy leads again to self-similar 
collapse \cite{Kagan} followed by intermittent 
implosion and pattern formation \cite{Pattern}, strikingly similar 
to a supernova implosion  \cite{Star}. On the contrary, for 1D BECs 
the attractive interactions are always compensated by quantum dispersion 
resulting in the formation of stable bright solitons \cite{Bright}. 

The partially attractive character of the DDI results in nontrivial
stability conditions for dipolar BECs when the DDI becomes sufficiently 
large compared to the short-range interactions. In particular, 3D homogeneous
dipolar BECs are unstable due to instability at very low momenta 
(phonon instability) \cite{Santos2000}. 
For trapped 3D dipolar BECs this instability leads to condensate collapse 
resembling the above mentioned case of short-range interacting gases, 
although the post-collapse dynamics is qualitatively 
different, being characterized by quadrupole-like patterns as recently 
observed experimentally \cite{CollapsePfau}. 

Due to the anisotropic character of the DDI, the 
gas instability is largely dependent on the trap geometry. 
Disc-shape traps surpassing a critical aspect ratio have been 
recently shown to prevent phonon-like instabilities 
if the dipoles are oriented orthogonal to the trap plane \cite{Stabilization}.
In that case, however, a sufficiently large dipole moment may lead to 
the rotonization of the Bogoliubov spectrum of the dipolar BEC \cite{Roton}, 
which eventually leads to local
collapses \cite{Komineas}, or stabilized modulated density profiles in
sufficiently small traps \cite{Ronen}.
 
Phonon instability in 2D traps is however present 
if the dipoles are oriented parallel to the trap plane.
The special long-range anisotropic character of the DDI 
leads naturally to the physically relevant question 
of how dimensionality affects the instability dynamics, and in particular 
whether 2D phonon instability is necessarily followed by condensate collapse.
In this Letter, we show that indeed 2D phonon instability in 
dipolar BECs does not necessarily lead to collapse, hence differing  
qualitatively from phonon instability in 2D 
and 3D short-range interacting BECs, and 3D dipolar ones.
The absence of collapse is explained by the formation of stable 
inelastic 2D bright solitons, whose stability results from the 
long-range character of the DDI \cite{Pedri2005,Yakimenko2006,Malomed},
being unstable in short-range interacting BECs. 
As a consequence, 2D phonon instability 
is accompained by the transient formation of a gas of 2D solitons, 
which resemble the formation of soliton trains in unstable BECs 
with $a<0$ \cite{Bright}. However, contrary to the latter case,  
2D dipolar solitons attract each other, and scatter inelastically 
\cite{Inelastic}, 
eventually fusing into larger solitons, which collapse only if the number 
of particles per soliton surpasses a critical value.
In the last part of this Letter, we show that in the presence of 
a significant in-plane harmonic confinement, 2D phonon-like instability 
leads to transient pattern formation followed by the creation of stable 
solitons. Interestingly, as we discuss at the end of this Letter, the 2D instability 
may be easily observed in on-going experiments, allowing for the creation 
of 2D stable solitons for the first time in quantum gases.




In the following, we consider a BEC of particles with mass $m$ and 
electric dipole $d$ (the results are equally valid for magnetic dipoles) 
oriented in the $z$-direction by a sufficiently large external field, and 
that hence interact via a dipole-dipole potential: 
$V_d(\vec{r})= \alpha d^2 (1-3\cos^2(\theta))/r^3$, where $\theta$ 
is the angle formed by the vector joining the interacting particles 
and the dipole interaction. The coefficient $\alpha$ can be tuned within 
the range $-1/2\leq\alpha\leq 1$ by rotating the external field that 
orients the dipoles much faster than any other relevant time scale 
in the system \cite{Tuning}. At sufficiently low temperatures the physics 
of the dipolar BEC is provided by a non-local non-linear Schr\"odinger 
equation (NLSE) of the form:
\begin{eqnarray}
&&i\hbar\frac{\partial}{\partial t}\Psi(\vec r,t)=\left [
-\frac{\hbar^2}{2m}\nabla^2+V(\vec r)
+g|\Psi(\vec r,t)|^2 \right\delimiter 0 \nonumber \\
&&+ \left\delimiter 0  \int d\vec r' V_d(\vec r-\vec r')
|\Psi(\vec r',t)|^2
\right ]\Psi(\vec r,t),
\label{GPE}
\end{eqnarray}
where $\int d\vec r |\psi(\vec r,t)|^2=N$, 
$g=4\pi\hbar^2a/m$, $a>0$ is the $s$-wave scattering length, 
$m$ is the particle mass and $V$ is the external 
confining potential. In the following 
we use the convenient dimensionless parameter $\beta=\alpha d^2/g$, 
that characterizes the strength of DDI compared to the short 
range interaction. Note that the ratio 
$\beta$ may be easily controlled experimentally by means of Feshbach 
resonances, as recently shown in Ref.~\cite{CollapsePfau}.




The 3D homogeneous ($V(\vec r)=0$) solution of Eq.(\ref{GPE}) 
is given by $\Psi(\vec r,t)=\sqrt{\bar n_{3d}}\exp{[-i\mu_{3d} t/\hbar]}$, 
where $\bar n_{3d}$ denotes the 3D density, and 
$\mu_{3d}=(g+\tilde V_d(0))\bar n_{3d}$ is the chemical potential, with 
$\tilde V_d(\vec q)=
(8\pi/3)\alpha d^2[3q_z^2/|\vec q|^2-1]/2$ the Fourier transform
of the DDI for dipoles oriented along the $z$-direction. 
A Bogoliubov analysis of the 3D homogeneous BEC 
provides the dispersion relation for quasiparticles: 
$\epsilon (\vec q)=[E_{kin}(\vec q)
[E_{kin}(\vec q)+E_{int}(\vec q)]]^{1/2}$, 
where $E_{kin}(\vec q)=(\hbar^2q_{\rho}^2+\hbar^2q_{z}^2)/2m$ 
is the kinetic energy, and $E_{int}(\vec q)=
2(g+\tilde V_d(\vec q))\bar n_{3d}$ 
is the interaction energy, which includes both 
short-range and dipolar parts. 
Note that $\tilde V_d(\vec q)$
may become negative for some particular directions, 
and hence for low momenta
phonon instability is just prevented 
if $-3/8\pi<\beta<3/4\pi$. If $\alpha>0$, phonons 
with $\vec q$ lying on the $xy$ plane are 
unstable if $\beta > 3/4\pi$, while for $\alpha<0$ 
phonons with $\vec q$ along $z$ 
are unstable if $\beta < -3/8\pi$. 
In both cases, the dipolar BEC is unstable against local collapses. 
In the presence of trapping, this instability leads to a global collapse of
the BEC, and may be geometrically prevented in sufficiently pancake-shapped
traps, as shown recently in Ref.~\cite{Stabilization}.




In the following we show that the phonon instability may become crucially different
in 2D dipolar condensates. We first consider the case of homogeneous 2D
dipolar BECs in the $xy$ plane, which are strongly confined by an harmonic 
potential $V=(1/2m)\omega_z^2 z^2$ along the transversal $z$-direction, such
that the system can be considered ``frozen'' into the ground 
state $\phi_0(z)$ of the harmonic oscillator in the $z$ direction, and  
hence the BEC wave function factorizes as  
$\Psi(\vec r)=\psi(\vec \rho)\phi_{0}(z)$. The dipole orientation with respect
to the trap plane plays of course a crucial role. In the following we discuss 
the two extremal configurations, in which the dipoles are normal to the $xy$
plane ($\perp$-configuration) and parallel to it (along $x$) ($||$-configuration).  
Employing the above mentioned factorization, the convolution theorem, the Fourier 
transform of the dipole-potential and integrating over 
the $z$ direction, we arrive at the 2D NLSE \cite{Pedri2005}:
\begin{eqnarray}
&&i\hbar\frac{\partial}{\partial t}\psi(\vec \rho)=
\left [ 
-\frac{\hbar^2}{2m}\nabla_{\rho}^2
+\frac{g}{\sqrt{2\pi}l_z}|\psi(\vec \rho)|^2 \right\delimiter 0 \nonumber \\
&+&\frac{4\sqrt{\pi}g \beta}{3\sqrt 2l_z} 
\left\delimiter 0  \int\frac{d\vec {k}}{(2\pi)^2}e^{i\vec
    k\cdot\vec {\rho}} 
\tilde n(\vec k)h_{2D}\left (\frac{ \vec {k} l_z}{\sqrt{2}}\right)
\right ]\psi(\vec \rho),
\label{GPE2D}
\end{eqnarray}
where $\vec k$ is the $xy$-momentum, 
$\tilde n$ the Fourier transform of $|\psi(\vec\rho)|^2$, 
$h_{2D}(\vec \kappa)=2-3\sqrt{\pi}\kappa e^{\kappa^2}$erfc$(\kappa)$ 
($\perp$-configuration), 
and  $h_{2D}(\vec \kappa)=-1+3\sqrt{\pi/2}(\kappa_x^2/k)
e^{\kappa^2}$erfc$(\kappa)$ 
($||$-configuration), 
with erfc$(\kappa)$ the complementary error function 
and $l_z=\sqrt{\hbar/m\omega_z}$.

The homogeneous solution of (\ref{GPE2D}) is 
$\psi(\vec\rho,t)=\sqrt{\bar n_{2d}}\exp{[-i\mu_{2d} t/\hbar]}$, 
where $\bar n_{2d}$ denotes the homogeneous 2D density, 
and $\mu_{2d}=g\bar n_{2d}(1+8\pi\beta/3)/\sqrt{2\pi}l_z$ is the chemical
potential for $\perp$-configuration and 
$\mu_{2d}=g\bar n_{2d}(1-4\pi\beta/3)/\sqrt{2\pi}l_z$ 
for $||$-configuration. The corresponding Bogoliubov 
equations provide the dispersion 
relation for elementary excitations of 2D BEC \cite{Yakimenko2006}:
\begin{eqnarray}
&&\epsilon(\vec k)^2 = E_{kin}^2+\frac{2gn_0E_{kin}}{\sqrt{2\pi}l_z}
\left [1+\frac{4\pi\beta}{3}h_{2D}\left(\frac{kl_z}{\sqrt 2}\right) \right]
\label{DISP1}
\end{eqnarray}
where
$E_{kin}=\hbar^2|\vec k|^2/2m$. 

We discuss first the $\perp$-configuration. From (\ref{DISP1}) 
we see that the excitations with low momenta ($\vec k\to 0$), 
become purely imaginary if $\beta < -3/8\pi$, i.e. phonon instability 
demands the tuning of the DDI, being absent without tuning. 
As discussed in Ref.~\cite{Pedri2005} stable 2D dipolar bright
solitons are possible for $g>0$ if $\beta<-3/8\pi$. In the following we call
these solitons {\em isotropic solitons} (since they are isotropic on the 
$xy$-plane) to distinguish them from the 
solitons introduced in Ref.~\cite{Malomed}, which we discuss below.
As observed, the phonon-instability and the soliton-stability
conditions are exactly identical. This fact, far from being accidental, 
shows that the phonon-instability and 2D soliton formation are
intrinsically linked. 

\begin{figure} 
\begin{center}
\includegraphics[width=4.8cm]{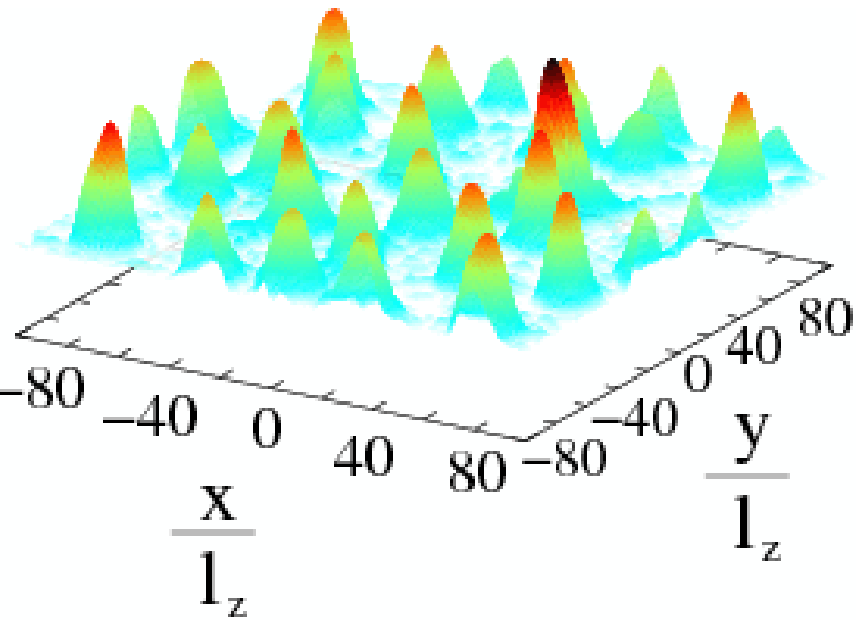}\\
\includegraphics[width=4.8cm]{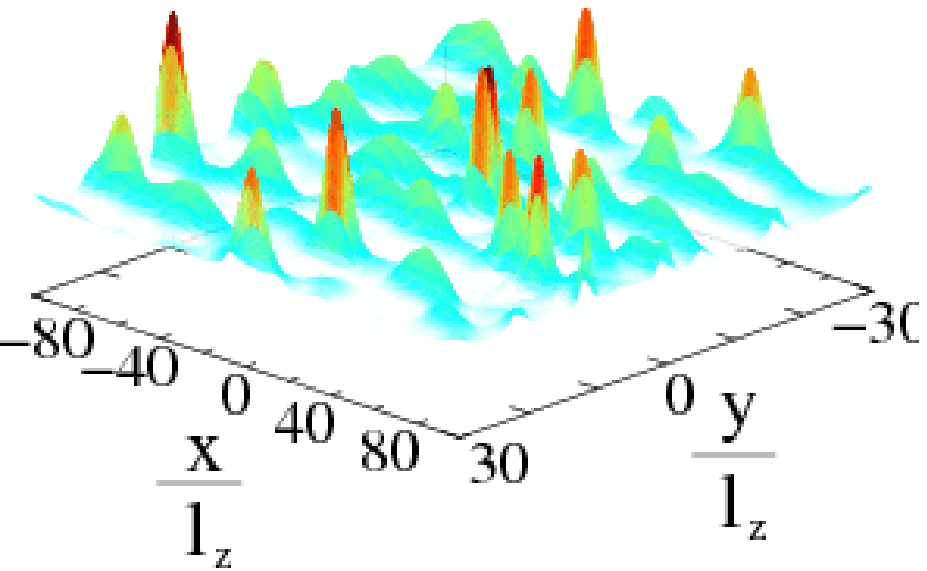}
\end{center}
\caption{2D soliton gas after phonon instability of an
  homogeneous dipolar BEC.  
For both the figures $\mu=-0.2$ and $g/(\sqrt{2\pi}l_z)=10$. 
(top) $\perp$-configuration, $\beta=-0.2$, $t=484/\omega_z$; \
(bottom) $||$-configuration, $\beta=0.3$, and $t=108/\omega_z$.}  
\label{fig:1}
\vspace*{-0.2cm}  
\end{figure}

Using Eq.~(\ref{GPE2D}), we have numerically studied the phonon 
instability by employing periodic boundary conditions. In all our simulations, we
check that $\tilde \mu_{2d}=\mu_{2d}/\hbar\omega_z< 1$ to ensure the 2D
character of the problem. Starting from an homogeneous solution, our real time 
simulations show the dynamical formation of a soliton gas, as shown in 
Fig.\ref{fig:1} (top). The long-range character of the DDI leads to an attractive 
soliton-soliton interaction (since $\beta<0$). As discussed in
Ref.~\cite{Inelastic} the soliton-soliton scattering is inelastic, and for 
low relative velocities (as it is the case in the soliton gas) soliton fusion
is expected. This is indeed observed, and hence the number of solitons 
decreases in time, whereas the number of particles per soliton increases.
As long as the problem remains 2D, the isotropic solitons remain stable
(contrary to the anisotropic solitons discussed below). However, the
density within a given soliton may eventually become large enough to violate the 2D 
condition $\tilde \mu_{2d}< 1$. In that case the soliton becomes unstable
against 3D collapse \cite{Pedri2005}.


\begin{figure} 
\begin{center}
\includegraphics[width=6cm]{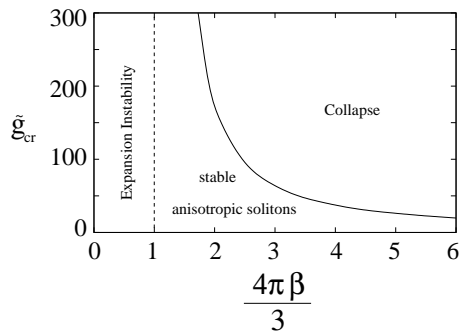}
\end{center} 
\vspace*{-0.2cm} 
\caption{Stability diagram of an anisotropic soliton as a function 
of $\beta$ and $\tilde g_{cr}=g N_{cr}/\sqrt{2\pi}l_z$, where for $N>N_{cr}$ 
the soliton is unstable against collapse even for $\beta>3/4\pi$.}
\label{fig:2}
\vspace*{-0.2cm}  
\end{figure}


The $||$-configuration shows a relatively similar physics.   
From Eq.~(\ref{DISP1}) we see that $\epsilon(\vec k\to 0)$ becomes unstable 
if $\beta>3/4\pi$. This condition is exactly that found recently in 
Ref.~\cite{Malomed} for the stability of 2D bright solitons. Following 
Ref.~\cite{Malomed} we call in the
following these solitons {\em anisotropic solitons}, since the solitons are
more elongated along the dipolar axis (in our case the $x$ axis). Before 
considering the instability dynamics in the $||$-configuration, we briefly 
comment on the properties of the anisotropic solitons, which differ 
to some extent from those of the isotropic solitons. As
mentioned in Ref.~\cite{Malomed} a simple Gaussian Ansatz on the $xy$-plane 
with unequal widths $L_{x,y}$, shows the appearance of a minimum 
in the energy functional $E(L_x,L_y)$ for some equlibrium soliton widths
$L_{x0}$ and $L_{y0}$. However, one may show  that 
for a given $\beta>3/4\pi$, there is a critical universal value 
$\tilde g_{cr} (\beta) \equiv g N_{cr}/\sqrt{2\pi}l_z$ (see Fig.~\ref{fig:2})
such that for $N>N_{cr}$ the minimum of $E(L_x,L_y)$ 
dissappears. As a
consequence, contrary to the case of isotropic solitons, which are always
stable as long as the soliton is 2D, there is a critical number of particles
per soliton, $N_{cr}$, which decreases for larger $\beta$. 
Beyond this number the 2D soliton collapses. This result is also 
verified by a direct simulation of the 3D Gross-Pitaevskii 
equation (\ref{GPE}).

Using Eq.~(\ref{GPE2D}), and starting once more from an 
homogeneous gas, we have 
observed the dynamical formation of a gas of anisotropic solitons, after 
initial formation of stripes along the dipole axis  
(Fig.~\ref{fig:1}, bottom).
Due to the anisotropic nature of DDI in the plane, the inelastic dynamics of
anisotropic solitons is relatively more complicated. 
Solitons along the $x$ axis attract each other, fuse together and 
become unstable against 2D collapse once they surpass the critical number of
particles per soliton.




In the final part of this Letter, we consider a 2D dipolar gas in the presence 
of an harmonic confinement on the $xy$-plane, $V(\rho)=m\omega_{\rho}^2\rho^2/2$. As an initial condition 
for our time evolution, we have considered a dipolar
BEC in the Thomas-Fermi regime (sufficiently large $g$) , 
obtained for the $\perp$-configuration with $\beta>0$ (no phonon instability) 
by imaginary time evolution of Eq.~(\ref{GPE2D}). In order to trigger the
phonon-like instability in our real time evolutions, at time $t=0$ we may
either switch $\beta<0$ (which keeps the polar symmetry)
or tilt the dipole switching into the $||$-configuration 
(hence violating the polar symmetry). 
We call these two cases, cases (i) and (ii).

\begin{figure} 
\begin{center}
\includegraphics[width=3.6cm]{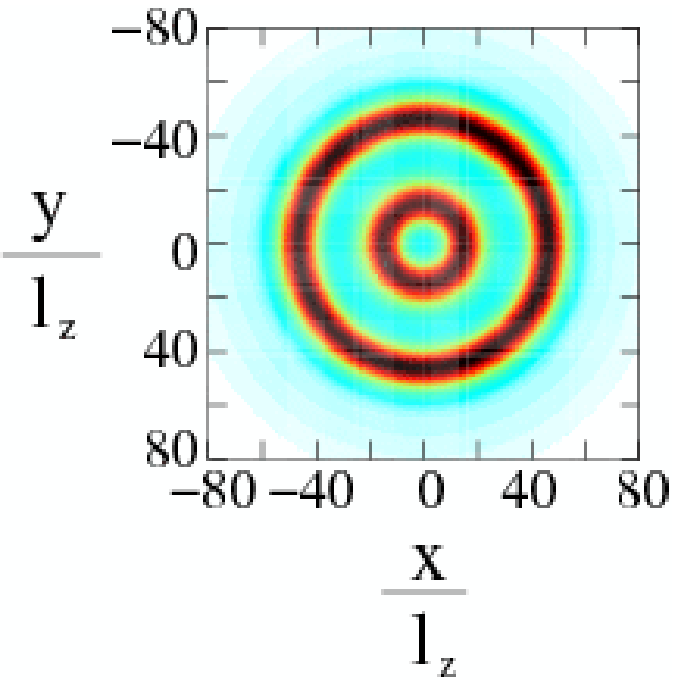}
\hspace{-0.1cm}
\includegraphics[width=3.6cm]{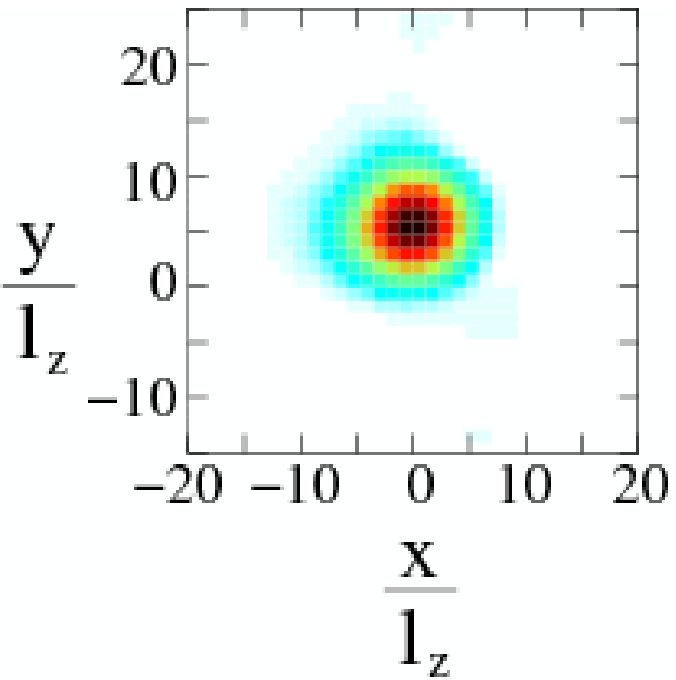}
\end{center} 
\vspace*{-0.3cm} 
\caption{Trapped BEC in the
  $\perp$-configuration. 
In the simulations, $\omega_{\rho}=0.001\omega_z$, 
$g/(\sqrt{2\pi}l_z)=200$ and $\beta=0.3$ ($t<0$) and $-0.3$ ($t>0$).
(left) Formation of ring structures ($t=810/\omega_z$); 
(right) Final soliton at the trap centre ($t=7500/\omega_z$).} 
\label{fig:3}
\end{figure}

Figs.~\ref{fig:3} illustrate the case (i). The
phonon-like instability leads to transient ring-shapped
structures. The harmonic confinement leads to successive
shrinkings and expansions of the rings, which oscillate in the trap.  
This process continues until the inner rings merge 
into a single excited dipolar bright soliton. We have observed that for
sufficiently shallow traps the ring structures develop azimuthal instability, 
which leads to 2D bright solitons, recovering, as expected, the 
homogeneous case.

\begin{figure} 
\begin{center}
\includegraphics[width=3.4cm]{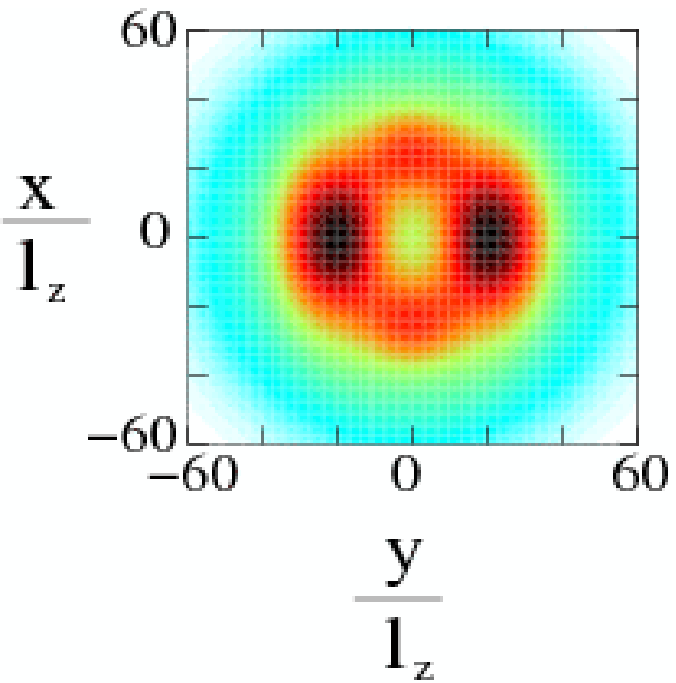}
\hspace{-0.1cm}
\includegraphics[width=3.4cm]{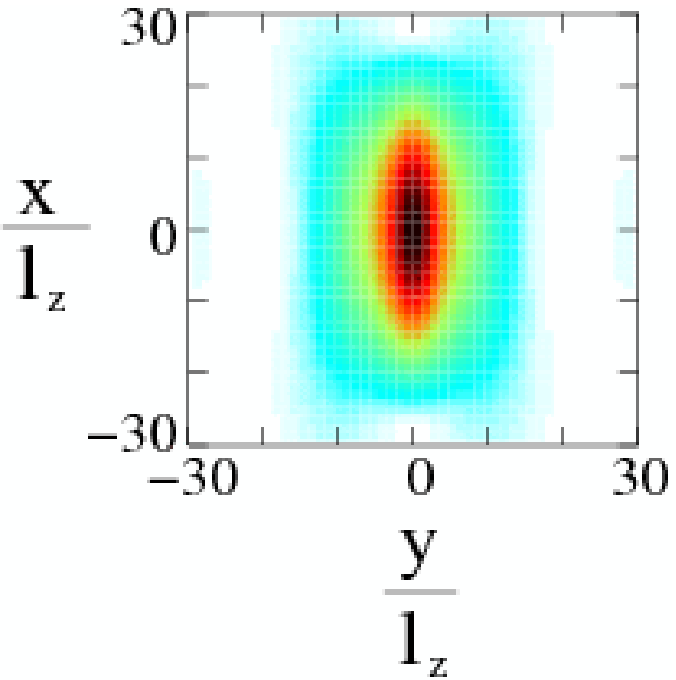}
\end{center} 
\vspace*{-0.3cm} 
\caption{Trapped BEC after tilting at $t=0$ 
from $\perp$ to $||$ configuration, with 
$\omega_{\rho}=0.001\omega_z$, $g/(\sqrt{2\pi}l_z)=200$ and $\beta=0.3$.
(left) Anisotropic ring formation ($t=1092/\omega_z$);
(right) Final formation of an anisotropic soliton ($t=1281/\omega_z$).}
\label{fig:4}
\vspace*{-0.2cm}  
\end{figure}

The case (ii) is illustrated in Figs.~\ref{fig:4}. An initial
anisotropic ring-like structure breaks into a pair of anisotropic clouds,
which eventually merge into the trap center forming a single  
stable bright soliton, which as expected is anisotropic, being more elongated
in the dipole $x$-direction. Note that both for the cases (i) and (ii) 
the non-dissipative character of the problem implies that the solitons 
are produced in an excited internal state.


Summaryzing, we have shown that the phonon instability of a 2D 
dipolar BEC differs qualitatively from 2D and 3D
short-range interacting gases, and 3D dipolar BECs. Contrary to
these cases, the phonon instability does not necessarily lead to the collapse 
of the gas, but on the contrary leads to a transient regime characterized by
the formation of a gas of attractive inelastic 2D bright solitons, which
eventually undergo fusion, leading to the creation of a 
single excited stable bright
soliton. If the dipoles are perpendicular to the trap plane these solitons 
are stable as long as the gas remains 2D, whereas if the dipoles are parallel
to the trap plane the (anisotropic) solitons may become unstable 
even in 2D for a critical number of particles per soliton. Finally, we have
shown that the instability in the presence of an harmonic confinement 
is followed by the creation of
transient ring-like and anisotropic patterns, which eventually lead to the
creation of a single excited 2D soliton. 

Finally, we comment on possible experimental realizations. 
Phonon instability for dipoles lying on the trap plane 
(and hence anisotropic solitons~\cite{Malomed}) does not demand the use 
of tuning, which may be experimentally demanding \cite{PfauPrivate}. 
Hence, 2D phonon instability should be relatively easy to 
observe in on-going experiments. A possible path
would be to 
prepare a 2D stable BEC orienting the dipoles perpendicular to the
trap plane, and then suddenly re-orient the dipoles on the plane. An 
equivalent alternative, more feasible experimentally \cite{PfauPrivate}, 
would be to maintain at any time the dipole on the trap plane, but drive the 
instability by increasing $\beta$ using Feshbach resonances as in 
Ref.~\cite{CollapsePfau}. As we have shown, 
the 2D instability would allow for the creation of 2D bright
solitons for the first time ever in quantum gases, since they 
are fundamentally impossible in absence of DDI. 2D solitons would 
offer new scenarios for the matter-wave 
soliton scattering, as e.g. 2D inelastic scattering and 
spiraling, similar to that in photorefractive materials 
\cite{Photorefractive}.

\acknowledgements
We acknowledge fruitful discussions with Th. Lahaye and T. Pfau. 
This work was supported by the DFG (SFB407, SPP1116). 
LPTMC is UMR 7600 of CNRS.

\end{document}